\pdfoutput=1
\documentclass[10pt,conference]{IEEEtran}
\usepackage{cite}
\usepackage{amsmath,amssymb,amsfonts}
\usepackage{algorithmic}
\usepackage{graphicx}
\usepackage{textcomp}
\usepackage{xcolor}
\usepackage[hyphens]{url}
\usepackage{fancyhdr}
\usepackage{hyperref}

\newcommand{\hpcayear}{2025}

\usepackage{graphicx}
\usepackage[space]{grffile}
\usepackage{soul}
\usepackage{booktabs}

\newcommand{\belowcaptionskipVALUE}[0]{0pt}
\usepackage[shortlabels]{enumitem}

\usepackage[normalem]{ulem}
\usepackage[ruled,linesnumbered]{algorithm2e}
\usepackage[utf8]{inputenc}

\usepackage{float}

\newcommand{\red}[1]{{\leavevmode\color{red}#1}}

\newcommand{\blue}[1]{{\leavevmode\color{blue}#1}}

\renewcommand{\red}[1]{#1}
\renewcommand{\blue}[1]{}

\newcommand{\hpcasubmissionnumber}{904}
\title{Skip TLB flushes for reused pages within mmap's}

\def\hpcacameraready{} 

\newcommand\hpcaauthors{Frederic Schimmelpfennig$\dagger$, André Brinkmann$\dagger$, Hossein Asadi$\ddagger$, Reza Salkhordeh$\dagger$}
\newcommand\hpcaaffiliation{Johannes Gutenberg University Mainz$\dagger$, Sharif University of Technology$\ddagger$}
\newcommand\hpcaemail{}

\author{
  \ifdefined\hpcacameraready
    \IEEEauthorblockN{\hpcaauthors{}}
      \IEEEauthorblockA{
        \hpcaaffiliation{} \\
        \hpcaemail{}
      }
  \else
    \IEEEauthorblockN{\normalsize{HPCA \hpcayear{} Submission
      \textbf{\#\hpcasubmissionnumber{}}} \\
      \IEEEauthorblockA{
        Confidential Draft \\
        Do NOT Distribute!!
      }
    }
  \fi 
}

\fancypagestyle{camerareadyfirstpage}{%
  \fancyhead{}
  
  \fancyhead[C]{
    \ifdefined\aeopen
    \parbox[][12mm][t]{13.5cm}{\hpcayear{} IEEE International Symposium on High-Performance Computer Architecture (HPCA)}    
    \else
      \ifdefined\aereviewed
      \parbox[][12mm][t]{13.5cm}{\hpcayear{} IEEE International Symposium on High-Performance Computer Architecture (HPCA)}
      \else
      \ifdefined\aereproduced
      \parbox[][12mm][t]{13.5cm}{\hpcayear{} IEEE International Symposium on High-Performance Computer Architecture (HPCA)}
      \else
      \parbox[][0mm][t]{13.5cm}{\hpcayear{} IEEE International Symposium on High-Performance Computer Architecture (HPCA)}
    \fi 
    \fi 
    \fi 
    \ifdefined\aeopen 
      \includegraphics[width=12mm,height=12mm]{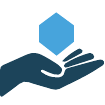}
    \fi 
    \ifdefined\aereviewed
      \includegraphics[width=12mm,height=12mm]{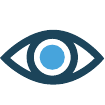}
    \fi 
    \ifdefined\aereproduced
      \includegraphics[width=12mm,height=12mm]{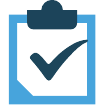}
    \fi
  }
  \fancyfoot[C]{}
}
\begin{document}
\maketitle


\newcommand{\hpcaheight}{0mm}
\ifdefined\eaopen
\renewcommand{\hpcaheight}{12mm}
\fi


\begin{abstract}

Memory access efficiency is significantly enhanced by caching recent address translations in the CPUs' \emph{Translation Lookaside Buffers} (TLBs). 
However, since the operating system is not aware of which core is using a particular mapping, it flushes TLB entries across all cores where the application runs whenever addresses are unmapped, ensuring security and consistency. 
These TLB flushes, known as \emph{TLB shootdowns}, are costly and create a performance and scalability bottleneck. 
A key contributor to TLB shootdowns is memory-mapped I/O, particularly during mmap-munmap cycles and page cache evictions. 
Often, the same physical pages are reassigned to the same process post-eviction, presenting an opportunity for the operating system to reduce the frequency of TLB shootdowns. We demonstrate, that by slightly extending the mmap function, TLB shootdowns for these "recycled pages" can be avoided.

Therefore we introduce and implement the "\textbf{f}ast \textbf{p}age \textbf{r}ecycling" (FPR) feature within the mmap system call. 
FPR-mmaps maintain security by only triggering TLB shootdowns when a page exits its recycling cycle and is allocated to a different process. 
To ensure consistency when FPR-mmap pointers are used, we made minor adjustments to virtual memory management to avoid the ABA problem. 
Unlike previous methods to mitigate shootdown effects, our approach does not require any hardware modifications and operates transparently within the existing Linux virtual memory framework.

Our evaluations across a variety of CPU, memory, and storage setups, including persistent memory and Optane SSDs, demonstrate that FPR delivers notable performance gains, with improvements of up to 28\% in real-world applications and 92\% in micro-benchmarks. 
Additionally, we show that TLB shootdowns are a significant source of bottlenecks, previously misattributed to other components of the Linux kernel.
        
\end{abstract}

\section{Introduction}
\label{sec:intro}

Leveraging parallel CPU resources results in significant performance improvements across a wide range of applications. 
This works well with advances in memory and storage solutions, such as persistent memory or Optane SSDs~\cite{EssenPAG12}, which have significantly reduced access times. However, as core counts increase and latencies decrease, operating system (OS) overhead becomes more pronounced~\cite{RadojkovicCVPGCNV08, WentzlaffA09}.

The management of Translation Lookaside Buffers (TLBs), e.g., has a significant impact on the performance of CPU architectures~\cite{CrottyLP22}. A TLB caches virtual-to-physical address mappings for its CPU core to reduce the number of page table walks. When a memory mapping becomes invalid, the OS removes that mapping from the TLBs to ensure security and consistency. Without any protection, a process could otherwise retain access to a physical memory location that it should no longer have access to, or reuse stale data.

The TLB is tightly embedded in the hardware, and its management by the operating system is rather coarse-grained. Flushes are limited to large portions or even the entire buffer, even if only individual entries become invalid. Therefore, the impact of a TLB flush includes not only the time for the flush operation itself, but also an indirect slowdown for rebuilding unnecessarily lost entries by falling back to the page table. Furthermore, the OS does not only flushes the TLB of the invalidating core, but also sends interrupts to all cores, that have \emph{possibly} accessed the address, to make them flush their TLBs. These \textit{TLB shootdowns} become more problematic as the number of cores used by an application increases.

Many shootdowns are associated with mapping memory for I/O, using \texttt{mmap()} and \texttt{munmap()}, or when the OS swaps out memory mapped pages during page cache eviction. 
The former is common in the Apache web server~\cite{HuNY99}, the latter happens within data\-bases and key-value stores such as LMDB~\cite{chu2020lightning}. LevelDB~\cite{ghemawat2014leveldb} mixes both access patterns.
During page cache evictions, the \emph{kswapd} process batches the least recently used pages and merges the CPU masks of its processes, resulting in shootdowns that target a high number of cores. 
TLB shootdowns are increasingly affecting the performance of those applications. 
The use of faster storage devices increases the rate at which data pages are swapped out and into memory, resulting in more shootdowns, especially for fast read accesses. 

\begin{figure}[b!]
  \centering
  \includegraphics[width=0.85\linewidth]{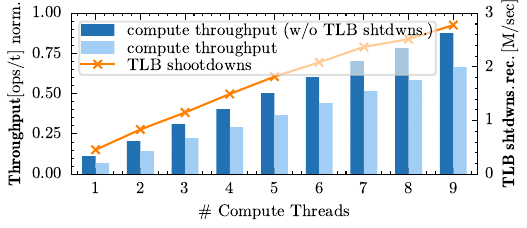}
  \caption{TLB shootdowns and compute performance for increasing number of compute threads and a single I/O thread.}
  \label{fig:intro-motivating}
\end{figure}

Figure~\ref{fig:intro-motivating} shows the impact that a single I/O thread performing mmap$\rightarrow$read$\rightarrow$munmap cycles has on independent compute threads of the same application. The compute threads do not access the data of the I/O thread. However, the OS considers the possibility that the compute threads might access pages being unmapped by the I/O thread. Therefore, all the compute threads are affected by the received TLB shootdowns initiated by the single I/O thread. This results in up to $30$\% waste of the overall compute throughput, with the absolute numbers increasing for higher thread counts.

Several previous approaches attempted to redesign the Linux memory management to reduce TLB shootdowns by identifying cases where one can be sure that an entry does not exist in a TLB, thus allowing shootdowns to be skipped ~\cite{Amit17, PapagiannisXSMB20, AwadBBSL17, VillaviejaKVERMNCU11, GugaleGMJ20, SkarlatosDGKT21, RomanescuLSB10}.
These approaches either add overhead to all TLB misses or required major changes to the Linux kernel. 
A second category of previous research focused on smaller optimizations of the Linux memory management, many of which are already part of the current mainline Linux kernel, such as batching~\cite{Uhlig05, PapagiannisMB21}, lazy shootdowns~\cite{ChangK97}, and overlapping local and remote flushes~\cite{AmitTW20}. 

Despite these efforts, TLB shootdowns continue to significantly impact the performance of data-intensive applications which use \texttt{mmap} for I/O. 
Furthermore, prior studies mainly focused on the impact of TLB shootdowns on I/O operations themselves, while TLB shootdowns caused by I/O operations also affect compute-intensive application threads.

The main insight used in this work is that the OS often recycles physical pages after \texttt{munmap()} or page cache evictions and reassigns them to the same application. 
From a security perspective, TLB shootdowns can be avoided when pages are reused by the same application, and TLB shootdowns only need to be sent when pages leave this recycling path. 
However, TLB shootdowns also ensure consistency, and we need to adjust the \texttt{mmap} internals to ensure that a new mapping assigned to the same virtual address as a previously unmapped one cannot be accessed by an old, invalid TLB entry, which would lead to the ABA problem~\cite{DechevPS10}.

We propose a new flag \texttt{MAP\_FPR} for the \texttt{mmap()} call that instructs the kernel to perform "\textbf{f}ast \textbf{p}age \textbf{r}ecycling" within user-defined mmaps, and present a lightweight implementation in the Linux kernel. FPR avoids TLB shootdowns without compromising security and consistency by tracking pages using an  \texttt{id} that uniquely represents a user-defined recycling environment. Possible contexts are a single flagged mmap, all flagged mmaps of a process, and custom setups. A shootdown is only sent when a page belonging to a flagged mmap leaves its recycling path. Both sources of TLB shootdowns, manual \texttt{munmap}s and page cache evictions, are targeted.

FPR is a lightweight extension to the current API. In our Linux implementation, FPR and unflagged, standard \texttt{mmap}s share the same page and memory management. Standard \texttt{mmap}s without our new flag work as usual based on their standard semantics. FPR can be selectively enabled for individual mappings, and developers can actively relieve their I/O bottlenecks. Optionally, we also provide an interception library that automatically adds the \texttt{MAP\_FPR} flag to all \texttt{mmap()} calls for user-defined I/O paths. This allows our new feature to be used without recompiling existing applications.

The FPR design ensures that no application can access data from another process, and that every correct program can directly use our interception library without changing any application properties other than performance. However, programs producing segmentation faults without FPR can access stale data when a core accesses a page that has been unmapped by another core. Although those accesses are limited to data within the program, FPR should only be used after an application has been successfully debugged and does not produce segmentation faults.

\red{We implemented FPR in recent Linux kernels and evaluate it using Intel and AMD multi-core machines with various configurations of memory and storage, such as Optane persistent memory (pmem), Optane NVMe SSDs, and traditional NVMe SSDs.}
Using real-world applications, we achieve performance improvements of up to 28\% (\texttt{munmap}), 4\% (eviction), and 12\% (a combination of both). 
We also developed a comprehensive set of micro-benchmarks covering different patterns of \texttt{mmap} I/O accesses combined with compute operations. 
Within micro benchmarks, we are able to improve throughput by up to 92\% for \texttt{munmap} and up to 8.5\% in eviction scenarios. 
Our benchmarks provide new insights into shootdown bottlenecks, particularly in the eviction path of the Linux kernel. 
Both the FPR and our benchmarks will be made publicly available.

Our main contributions are as follows:
\begin{enumerate}[noitemsep,leftmargin=*,topsep=0pt]
    \item The introduction of FPR to effectively eliminate \texttt{mmap} TLB shootdowns for cases of page recycling during mmap.
    \item A benchmark suite on the impact of TLB shootdowns for various \texttt{mmap} usage patterns, allowing for specific optimizations of scalability bottlenecks.
    \item The evaluation of FPR using benchmarks and real-world applications on recent hardware configurations.
\end{enumerate}

The remainder of this paper is structured as follows. Section \ref{sec:background} presents the background. The related work is discussed in Section \ref{sec:related_work}. The concept of FPR is introduced in Section \ref{sec:design} and evaluated in Section~\ref{sec:evaluation}. Finally, Section~\ref{sec:conclusion} concludes the paper.
\section{Background}
\label{sec:background}

In this section, we provide an overview of memory management, \texttt{mmap()}-operations, and TLB shootdowns in the Linux kernel for the x86 architecture. 
Next, we present the Linux memory allocation mechanism and highlight the management of pages, which we use in our FPR-approach to minimize the number of shootdowns.

\subsection{Memory management and mmap}

Processes perform memory operations using virtual addresses that must be translated into physical addresses before the CPU can access them. 
The virtual-to-physical mappings, which are specific to each process, are stored in a page table in main memory. 
In the x86 processor architecture, the operating system manages the page table entries, while the CPU cores use these entries to translate virtual addresses to physical addresses through hardware page table walks. 
If the corresponding memory page for a virtual address does not exist in the page table, a \textit{page fault} is raised. 
The operating system then allocates a memory page, loads its contents based on existing memory mappings, and updates the page table. 
The CPU then retries the address translation. 
\blue{Getting a mapping requires multiple memory accesses, which can degrade performance.}

The \texttt{mmap()} syscall adds a mapping to the virtual address space of a process. Applications use \textit{anonymous} page mappings to allocate volatile memory. \texttt{mmap()} can also be used to map a range of virtual addresses to the contents of a file or (block) device. File mappings allow an application to operate on the mapped file using memory operations to change file contents. The mapped file can be larger than the available memory, and memory management operations like swapping and syncing are used by the OS to provide a transparent view of the file. Users can choose between two mapping modes. \textit{Shared} mappings keep the state of the file consistent and \textit{private} mappings do not persist modifications back to persistent storage.

Memory mappings can be removed via \texttt{munmap()} and dirty pages are flushed to storage. 
Finally, the allocated memory pages are freed. 
Memory pages are also removed by the \textit{kswapd} daemon in case of memory pressure. 
\textit{kswapd} iterates over the least recently used pages and releases batches of 32 evictable, e.g., non-anonymous pages. 
This is known as page eviction or reclamation.  

\subsection{The Translation Lookaside Buffer (TLB)}

\begin{figure}[t]
  \centering
  \includegraphics[width=0.82\linewidth]{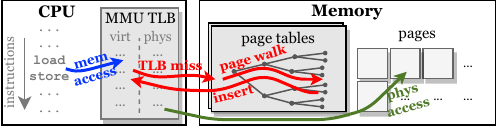}
  \caption{Interaction of the TLB and memory.}
  \label{fig:background-TLB}
\end{figure}

Most CPU architectures cache virtual-to-physical mappings in a hardware \emph{Translation Lookaside Buffer} (TLB) within the \emph{Memory Management Unit} (MMU) to prevent frequent and costly page table walks. Each physical core has a dedicated TLB that can hold up to 2,048 entries (x86)~\cite{EversBC22}. All memory accesses pass through the TLB and only fall back to a page table walk in case of a TLB miss (see Figure~\ref{fig:background-TLB}). When the CPU accesses a memory location, it queries the TLB to translate the virtual address into the physical address. If the mapping is not found (TLB miss), it traverses the process page table and inserts the mapping into the TLB.

The operating system removes invalid TLB entries via \emph{interprocessor interrupts} (IPIs). 
Invalidations sent to all cores running an application are called \textit{TLB shootdowns} and take a significant amount of time to process.
Most processor architectures, including x86, support multiple TLB flush granularities~\cite{IntelManual}.
Full TLB flushes are fast, but also remove valid entries, reduce the TLB hit ratio, and require reinsertion of unnecessarily lost entries.
Restricted address range flushes, on the other hand, are slower per IPI. 

TLB shootdowns are triggered by the operating system when a \texttt{munmap()} system call removes a page mapping or when a page is evicted from the page cache by the \textit{kswapd} daemon. 
Additionally, the OS sends a shootdown if a page is migrated between different NUMA nodes and it, therefore, changes its physical address. 

Linux uses several optimizations to reduce the shootdown overhead. 
First, it tries to batch requests to reduce the number of shootdowns, since the overhead of a shootdown to remove a single entry is almost the same as removing many entries. 
Linux also keeps a bitmap of the CPU cores running each application, and only sends shootdowns for an application to cores that are running it. 

\begin{figure}[!b]
  \centering
  \includegraphics[width=.95\linewidth]{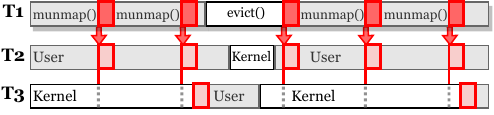}
  \caption{Varying impact of TLB shootdowns from thread T1 for T2, mostly in user space, and T3 mostly in kernel space.}
  \label{fig:TLB-impact-usage-patterns}
\end{figure}

Linux also \emph{lazily} delays shootdowns when the core is running a kernel task. 
The core then does not access user addresses directly, so invalid TLB entries are not used. 
When returning to user space, all received shootdowns are processed together. 
Fig.~\ref{fig:TLB-impact-usage-patterns} shows a thread T1 causing shootdowns, with threads T2 and T3 receiving them. 
T2, which runs mostly in user space, is more affected by incoming shootdowns than Process T3, which runs mostly in the kernel.

The x86 architecture supports \emph{process context identifiers} (PCIDs), which allow a subset of tagged TLB entries to be used for address translation~\cite{IntelManual}.
In Linux, they are used to distinguish between user and kernel space entries, e.g. in the context of some KAISER/Meltdown fixes. However, PCIDs are limited to 4096 values, and they cannot help for shootdowns within a process.

\subsection{Memory allocation}

Linux uses a global \textit{buddy allocator} to handle free memory \cite{Knowlton65}. 
A buddy allocator partitions free memory into $n$ lists $\{0, 1, ..., n-1\}$ and all free memory blocks of size $2^i \cdot m$ bytes are stored in list $i$. 
$m$ is the minimum size that can be allocated through the buddy allocator, and is typically set to the page size of the processor (e.g., 4 kB for x86 processors), while $2^{n-1} \cdot m$ is the maximum size of contiguous memory that can be allocated.  

Linux tries to fulfill an allocation request by picking a free block from the best-fitting list. If no such block exists, the next larger free block is recursively split or smaller blocks are collected, until they can serve the allocation. Splitting a block leads to two smaller blocks of the same size which are called buddies. If a block is freed and its buddy is also free, they are (recursively) merged back into bigger free memory blocks, before they are added to the according \textit{free list}. 
The buddy allocator uses locking that limits scalability.

Each CPU, therefore, maintains an own list of free pages used for paged-sized memory requests in a \textit{fast path} without acquiring a lock. If a CPU's list becomes empty, a large chunk of pages is allocated through the buddy allocator (slow path). If the global buddy allocator cannot serve the requests, pages will be removed from other CPUs' lists. Many applications constantly allocate and free pages from their CPU lists. In the context of memory mappings, this means that \texttt{mmap}-\texttt{munmap} cycles and re-allocations following page cache evictions mostly use the same set of physical pages. In this paper, we will exploit this re-using of pages to reduce the number of TLB shootdowns.
\section{Related Work}
\label{sec:related_work}

\blue{In this section, we present related work on approaches to prevent shootdowns and to mitigate their impact.} 

\subsection{Hardware approaches}

Self-invalidating TLB entries (SITE) add an expiration time to each TLB entry and expired entries are considered as invalid by the CPU~\cite{AwadBBSL17}. 
The OS does not need to send shootdowns for such expired TLB entries, reducing shootdowns by up to 65\%. 
DiDi includes hardware bookkeeping of all insertions and deletions of TLB entries to reduce the number of cores receiving shootdowns~\cite{VillaviejaKVERMNCU11}. 
ATTC maintains an inverse mapping table to eliminate shootdowns with minimal hardware changes in virtualized environments~\cite{GugaleGMJ20}. 
HATRIC uses cache coherence protocols to piggyback information during page migrations to reduce the number of TLB shootdowns and also targets virtualized environments~\cite{YanVCB17}. 
Mosaic Pages use compressed TLB entries to increase the overall TLB range~\cite{GosakanHKMMSTWB23}.
The presented HW approaches can reduce the performance impact of TLB shootdowns but require (significant) changes to current processor architectures.

\subsection{Software approaches}

The impact of TLB shootdowns can be decreased if multiple shootdowns are batched into a single inter-processor interrupt (IPI), either through the syscall interface~\cite{Uhlig05, Uhlig07} or directly in the kernel~\cite{AmitTW20}. 
Furthermore, shootdowns can be batched and performed lazily if the memory consistency protocol is relaxed and an additional communication channel is introduced~\cite{ChangK97}.

Several software approaches reduce the number of TLB shootdowns by tracking pages that are accessed from more than one CPU core and only send shootdowns for such pages. 
The Access-Based Invalidation System (ABIS) uses the access bit of page table entries to track whether more than one CPU core accesses a page~\cite{Amit17}. 
It therefore temporarily changes the \texttt{cr3} register to a fake page table to load the address mapping into the first accessing core. 
Then it uses the access bit to detect a second CPU core accessing it. 
The tracking needs to be enabled globally and imposes up to 9\% general performance overhead, whereas it can improve performance by up to 78\% for multi-core applications. 
Unfortunately, the implementation is not compatible with newer Linux kernels and requires heavy code updates. 
z-READ~\cite{ParkMYS19} has employed a similar approach to accelerate read operations and to avoid the copying of data.

Amit el. al have identified several performance limitations in the TLB shootdown path and have shown that aggressive batching can lead to correctness issues~\cite{AmitTW20}. 
They instead optimized the time for TLB shootdowns by improving the underlying protocols. 
These optimizations are orthogonal to our proposed architecture, which reduces the number of TLB shootdowns. 
They also identified specific situations like copy-on-write, where shootdowns can be safely removed. 
The underlying code is unfortunately not available. 

LATR~\cite{KumarMKVYKBK18} avoids reusing virtual addresses so that TLB shootdowns can be prevented. 
TLB invalidation is then performed during context switches by the core itself, thus, removing the need for expensive IPIs. 
Unfortunately, the kernel has no control over the usage of virtual addresses during page eviction and LATR does not cover this scenario. 
The underlying idea has been used in~\cite{AmitTW20} to batch shootdowns while the \texttt{mmap} semaphore is acquired, as virtual address cannot be re-used during this time period. 
EcoTLB makes LATR compatible with the way Linux uses PCIDs for the \emph{lazy} shootdown optimisations and to support disaggregated memory~\cite{MaassKKKB20}.

Fastmap~\cite{PapagiannisXSMB20} addresses performance and scalability problems of page eviction by re-designing \texttt{mmap}. 
A part of their improvements reduces the number of TLB shootdowns. However, their implementation requires a pre-allocated memory region and cannot operate within the current Linux memory subsystem.  
Additionally, it breaks the current \texttt{mmap} interface and cannot be used transparently. 
To further investigate the impact of TLB shootdowns on \texttt{mmap}, we also compare Fastmap with our scheme.
\section{Fast page recycling}
\label{sec:design}

In this section, we introduce FPR and how it reduces the number of TLB shootdowns for data-intensive applications that employ \texttt{mmap()}. FPR is an extension to the standard \texttt{mmap} that is enabled for specific mappings by passing the new flag \texttt{MAP\_FPR} to a \texttt{mmap()} call. It integrates seamlessly with the default \texttt{mmap}. 

The main idea behind FPR is that shootdowns can be delayed if a page is recycled. Recycling means that a page is reused in the same context that unmapped it before. We therefore define recycling as a passive state that describes the reuse of the physical page. Every page can exit the recycling state at any time due to an allocation to a different mapping (e.g., a default non-FPR mapping or an unrelated FPR mapping). Also, the kernel does not actively force pages within FPR mappings to be recycled.

We will show in the evaluation section that FPR can eliminate nearly all TLB shootdowns for \texttt{mmap()}-heavy applications that perform read or copy-on-write operations. FPR ensures the following security and consistency guarantees:

\begin{enumerate}[noitemsep,topsep=0pt,leftmargin=*]
    \item Security: Processes can never access a stale TLB entry for a page after a TLB flush has been skipped if this page has left the recycling path and has been reallocated to another process.
    \item Consistency: Applications cannot access a stale page entry in the TLB after an unmap operation for an FPR-enabled mapping if the same access would not have produced a segmentation fault without using FPR.
\end{enumerate}

Security is achieved by tracking pages and sending a shootdown if a page leaves the recycling path and consistency is achieved by extending the virtual address assignment and \emph{kswapd} changes.

\subsection{Security: Tracking page recyclings}
\label{sec:design:tracking}

We have modified the kernel so that the operating system assigns an additional ID to each physical page frame that represents a recycling context. This ID is retained even when a page is returned to the allocator's empty page pool. The IDs are initially set to zero, indicating that no recycling is expected. If a page is allocated for any non-FPR use, its ID is set to zero. Only the IDs of pages allocated by an FPR-flagged mapping are set to their recycling context.

\begin{figure}[t!]
\centering
\includegraphics[width=.85\columnwidth]{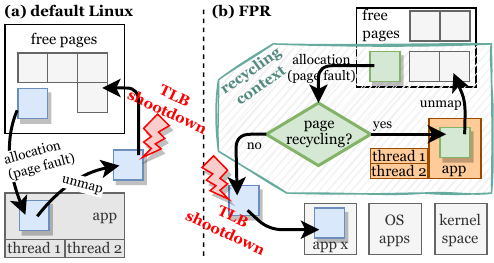}
\caption{FPR: Recycling pages are reused without shootdowns. They still are managed by the default OS internals.}
\label{fig:design}
\end{figure}

When a page is unmapped by \texttt{munmap}, the TLB shootdown is skipped if its ID is not equal to zero, indicating a page that is expected to be recycled within the same context again. This allows us to avoid unnecessary TLB shootdowns when pages are continuously recycled. For these pages, we move the shootdown logic from the release phase to the allocation phase. 

A change of the ID from a non-zero value during allocation indicates that the page has left a particular recycling context. In this case, a TLB shootdown is initiated to prevent erroneous memory accesses. This also means that for non-FPR pages, the default shootdown process still applies during the release phase. Fig.~\ref{fig:design} compares standard Linux page mapping with FPR, which prevents TLB shootdowns in \texttt{mmap} page recycling scenarios.  

\subsection{Consistency of applications using FPR}
\label{sec:design:vachange}

\begin{figure}[t]
  \centering
  \includegraphics[width=0.83\linewidth]{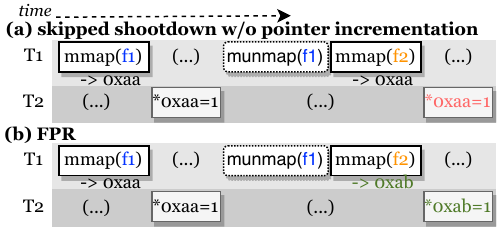}
  \caption{Usage of stale TLB entries during mmap-munmap cycles and FPR's virtual address iteration.}
  \label{fig:design:aba}
\end{figure}

Sending shootdowns when pages are leaving their recycling contexts ensures security but cannot guarantee the consistent execution of a process at the same time. 
Linux reassigns previously used virtual addresses during both, mmap-munmap cycles or through the page cache. 
Fig.~\ref{fig:design:aba} (a) shows an example, where thread T1 first frees a page for mapping \texttt{f1} and Linux then reallocates a different physical page to the same virtual address at the next \texttt{mmap}-call of the same thread T1 for mapping \texttt{f2}. 
In this case, FPR would not have sent any shootdowns to other cores running the same process. Now let us assume that the threads T1 and T2 concurrently access the same mappings. 
In this case, T2 would not update its corresponding TLB when accessing \texttt{f2} for the first time, as no page fault is produced based on the valid TLB entry and T2 would access the wrong physical page. 
This so-called ABA-problem is then leading to consistency errors~\cite{DechevPS10, StamlerHRZP22}. 

FPR therefore extends the virtual memory management and increments the virtual addresses that the kernel chooses for new mappings, instead of returning the same ones repeatedly (see Fig.~\ref{fig:design:aba} b). 
This is achieved by remembering the last assigned virtual address per process, so that searches for free virtual address ranges do not start from the lowest address. 
In the special case that the user forces a virtual address to be used by \texttt{mmap}, a shootdown is sent to comply with the request without changing the behavior of the system.

There is exactly one situation where the semantics of FPR-based mmaps differ from the standard mmap implementation. Again, suppose that two processes T1 and T2, running on different cores, concurrently access a mapping \texttt{f1}, and then \texttt{f1} is unmapped by T1. If T2 accesses \texttt{f1} after the unmap operation, this would cause a segmentation error in the standard implementation. However, when using FPR-based mmaps, T2 would be accessing an out-of-date physical page, which could lead to silent data corruption. Therefore, it is safe to add the \texttt{MAP\_FPR} flag to all \texttt{mmap} calls after successfully debugging an application \textbf{but} care must be taken as long as the application might throw segmentation errors.

\texttt{kswapd} requires a different handling because TLB shootdowns for pages evicted by \texttt{kswapd} cannot be skipped until a recycling-ending allocation, and because the virtual addresses can only be incremented within an \texttt{mmap}-call.  We therefore avoid shootdowns for pages getting recycled before an explicit \texttt{munmap} by lowering the eviction frequency and increasing the batch size for pages in a recycling context. Therefore, we do not evict FPR pages in a recycling context in case that \texttt{kswapd} works between the \emph{low} and \emph{high} memory watermark. Instead, we evict FPR pages in a recycling context only when the free memory reaches the \emph{min} watermark. In this case, we construct a huge batch of pages to be evicted, to free memory back to the high watermark. For this huge evict, a single shootdown must be sent.

\subsection{Design and implementation details}

The following subsection discusses further design choices and implementation details.

\subsubsection{Targeted use cases}

FPR optimizes fast and frequent I/O accesses through mmap, both mmap-munmap cycles and evictions of larger mappings. 
Especially read accesses tend to be impacted by TLB shootdowns, as there is no costly synchronisation or write-back, that could be executed concurrently to the TLB flushes.

FPR also supports anonymous memory and private mmap writes. However, FPR does not further optimize performance for anonymous mmap-munmap cycles (essentially cycles of \emph{malloc} and \emph{free}), because the \texttt{malloc} implementation in libc already includes optimized user-space caches. Also, private mmap writes at frequencies as high as in read scenarios are rare due to the underlying copy-on-write mechanisms.

\subsubsection{Extended recycling contexts}

In this paper, we only discussed recycling contexts for a single process. However, FPR also supports larger user-defined contexts:

\begin{enumerate}[noitemsep,topsep=0pt,leftmargin=*]
    \item One recycling context per FPR mmap ($tracking\_ID=(pid<<sizeof(mmap\_id))+mmap\_id$). Recycling is restricted to page cache evictions, as mmap-munmap are using multiple consecutive mappings.
    \item One recycling context for all FPR mmaps of a process, so that $tracking\_ID=pid$.
    \item One recycling context for all FPR mmaps of a process' child processes ($tracking\_ID=parent\_pid$). Enables FPR for shared mappings within the child processes.
    \item One recycling context for all FPR mmaps of all processes of a user ($tracking\_ID=uid$). Enables FPR for shared mappings within all processes of the user.
\end{enumerate}

The latter two settings require the user to trust the processes, as extending the context of an FPR mapping beyond a single process allows a process to create stale entries to physical pages which can be used to access data from other processes. 

\subsubsection{Enabling FPR}

We designed FPR as an opt-in solution using the \texttt{MAP\_FPR} flag, because the standard mapping is required to catch and debug segmentation faults. 
Apart from the basic way of using FPR via the \texttt{MAP\_FPR} flag, we provide an optional interception library that adds the flag automatically to all \texttt{mmap()} calls for user-defined paths (e.g., the database of a learning workflow). 
This way, existing applications can transparently make use of FPR without recompilation.

\subsubsection{Interaction with other sources of shootdowns}

Memory mappings can also be modified by other parts of the kernel. Page migration, e.g., physically moves pages and modifies their page table entries~\cite{RyooJB18}. Page migration can be explicitly requested or is triggered automatically by the \texttt{kswapd} process. We handle both cases in the same way and treat migration similarly to reallocating the migration target page. If the target page leaves a recycling context, we perform the shootdown. If the source page is within a recycling, we copy its tracking data to the new page. Freeing the source page is handled by \texttt{kswapd}, similar to evicted pages.

We need to track recycling pages in all places they are managed, and hence, we need to also track them if they are freed and moved back from the per CPU page lists to the buddy allocator. 
If two buddies are merged and one of them is tracked as recycling, we also track the merged buddy in the same recycling context. 
If both of them are tracked, but with different IDs, we set the second flag in our tracking data to state that a shootdown should be always sent for this memory region. 
The value of the version is set to the maximum of the former buddies. 
When the buddy allocator performs a split, we copy the tracking data to the split buddies. 
\blue{However, if a full shootdown was issued for the original buddy due to an recycling-leaving allocation, there is no need to copy the tracking data, as the stale entries have already been cleared.
}

\subsubsection{Further optimizations}

TLB shootdowns can be merged when several pages leave recycling. 
We, therefore, add a global counter, which is incremented for each global shootdown. 
The current value of this counter is set in the version field of our tracking data when a recycled page is freed. 
During a following reallocation, we only send a shootdown if the page is leaving recycling \textit{and} its version value matches the global counter. 
Otherwise, a global shootdown before freeing this page has already removed \textit{all} invalid entries from the TLBs. 
This eliminates the necessity of requiring shootdowns for each single page leaving a recycling.

When a process is forked, no additional operations are performed if the forked mappings stay within the recycling context. If this is not the case, a single TLB shootdown for the corresponding pages can be issued, before the forked processes are considered as new recycling contexts. Additionally, NUMA-rebalancing using FPR is supported by checks during the allocation of the target page-frame and future reallocation of the old source page-frame.

\subsubsection{Kernel implementation and memory footprint}

The changes made to the kernel are lightweight. Most functions are outsourced into a single header-file ($\sim$400LOC), to be used and inlined within \texttt{filemap.c}, \texttt{memory.c}, \texttt{mmap.c}, \texttt{mmu\_gather.c}, \texttt{page\_alloc.c} and \texttt{vmscan.c}. Within these files, a total of $\sim$60LOCs are modified. 
The kernel interfaces remain unchanged.

For each data page, we maintain 8 bytes (0.5\% overhead for 4 kByte pages) to track pages. 2 bits are used for flags, 22 bits for the ID, and 40 bits for the version. We store the tracking data in the same NUMA node as the corresponding pages. 
\section{Evaluation}
\label{sec:evaluation}

This section evaluates FPR in \texttt{munmap} and eviction scenarios using microbenchmarks and real-world applications. 
We also discuss how FPR performs in comparison with previous approaches.

\subsubsection{Experimental setup}

The overhead of TLB shootdowns depends on the CPU and therefore, we have evaluated FPR on different hardware configurations and employed three Intel- and AMD-based servers (see Table~\ref{table:eva.setup}). 
Two servers are dual-socket to investigate the impact of inter-socket communication on shootdowns. 
The single-socket Intel server is equipped with persistent memory to evaluate the impact of shootdowns for very fast storage.

\red{We have implemented FPR in longterm Linux kernel 5.15 and also ported it to to the most recent kernel 6.10.
Unfortunately, not all servers were available to us at this time, so we are showing the results for 5.15, unless otherwise mentioned.
Compared to the kernel 6.10, FPR's performance improvements have remained the same in terms of quality.}

We use the \emph{null block device driver} (\emph{nullblk}) in some experiments to avoid any device latency. 
As \textit{nullblk} cannot serve real-world applications, we additionally use the \textit{EXT4} file system that has been mounted on a RAM disk. 
\blue{\textit{EXT4} has the advantage, e.g., over \emph{tmpfs}, that the available page cache is also used by the OS.} 
\blue{André: This sounds as we have not used SSDs at all.}

\begin{table}[b!]
\centering
\caption{Hardware setup}
\label{table:eva.setup}
\begin{tabular}{@{}cccc@{}}
Server & CPU & Memory  & Storage \\ \midrule
\textit{Intel} & \begin{tabular}[c]{@{}r@{}}Intel Xeon 9242 \\ 96 cores, 2 socket\end{tabular} & 768G & \begin{tabular}[c]{@{}r@{}}Intel P4511 NVMe 2T\\ Optane P4800X 375G\end{tabular} \\ \hline
\textit{AMD} & \begin{tabular}[c]{@{}r@{}}AMD Epyc 7413 \\ 48 cores, 2 socket\end{tabular} & 256G & Toshiba PM5 SAS 1T  \\ \hline
\textit{pmem} & \begin{tabular}[c]{@{}r@{}}Intel Xeon 5215 \\ 10 cores, 1 socket\end{tabular} & 192G & Optane DC NVMM 768G 
\end{tabular}
\end{table}

\subsubsection{Methodology}

Pinpointing the overhead of shootdowns to a fixed location is not possible, since part of the overhead belongs to the interrupt handling routine that spans over hardware and software. Additionally, shootdowns can cause the eviction of seemingly unrelated TLB entries, which might increase the latency of other parts of an application. Therefore, we used the end-to-end runtime and throughput as evaluation metrics to consider all shootdowns effects. 

We cleared all relevant caches and copied the source data from a predefined image to exactly the same location before each test to remove the side-effects of previous runs. 
The memory usage and NUMA allocation strategies are the same across all experiments. 
We also cleared our tracking data and flushed TLBs and the page cache before each experiment when evaluating FPR. 

The standard deviation of our experiments is marginally small, and hence, omitted from the figures. 
All runs are performed at least three times for a sufficient duration (up to 2h for the page eviction cases) to obtain stable results. 
We report the average values of repeated experiments. 
FPR runs were performed with all security and consistency features enabled. 
The number of TLB shootdowns in the results denotes the sum of the remote TLB flush requests \emph{received} and executed.

Previous studies did not include scenarios to specifically analyze the impact of shootdowns on compute phases or on non-I/O threads~\cite{Amit17, CrottyLP22}. 
Some studies also claimed that the page eviction algorithm is slow and TLB shootdowns during page eviction would not impact application performance~\cite{Amit17, PapagiannisXSMB20}. 
In contrast, our microbenchmarks also consider shootdowns sent from a group of I/O threads to a set of compute threads and investigate cases where all threads perform a mixture of I/O and compute. 
Our microbenchmarks include both \texttt{munmap} and page eviction scenarios. 
In addition to the microbenchmarks, we have also evaluated FPR using the real-world applications Apache web server, LevelDB, and LMDB.

\red{The oldest kernel version running on our hardware is Linux 4.14. We therefore cannot provide a direct comparison of FPR with previous studies, which are only available for older Linux kernels (e.g., 4.5~\cite{Amit17} or 4.10~\cite{KumarMKVYKBK18, MaassKKKB20}). 
Porting their implementations to newer kernels requires a significant engineering effort, since the Linux kernel included several major changes in recent years and comparing our results with the published results using different setups would not take into account changes in the Linux kernel affecting the base results. 
Therefore, part of the comparison must be indirect.
Fastmap~\cite{PapagiannisXSMB20} is the only approach that runs on kernel version 4.14 and we, therefore, ported back FPR to it to provide a comparison with Fastmap. 
}

\subsection{mmap/munmap}

In this section, we evaluate the performance of our approach for \texttt{mmap/munmap}-heavy applications. 
First, we measured the performance using several microbenchmarks to show the effect of our optimizations. 
Next, we evaluated FPR using real-world applications that heavily use \texttt{mmap/munmap}.

\subsubsection{Microbenchmarks}

\begin{figure}[t]
  \centering
  \includegraphics[width=0.95\linewidth]{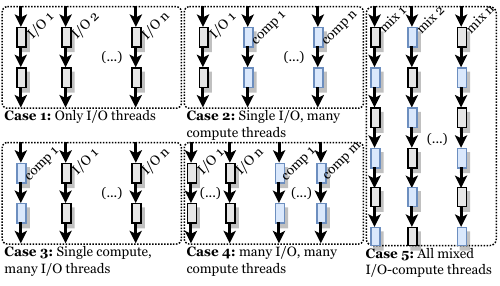}
  \caption{\textit{mmap}-\textit{munmap} cases overview.}
  \label{fig:eva.mmap-munmap.case-overview}
\end{figure}

Our benchmarks include three types of threads: 1) I/O threads performing loops of \textit{mmap}->\textit{4k-access}->\textit{munmap}, 2) compute threads continuously executing an instruction accessing a register to simulate compute-heavy functions, and 3) \textit{mixed} threads which alternate performing I/O and compute to simulate a process accessing data and then processing it. Since the results for the Intel and AMD servers are similar, we only report the results for the Intel server. We designed five combinations of these thread types and evaluated them on a shared mapped \textit{nullblk} device. Fig.~\ref{fig:eva.mmap-munmap.case-overview} provides an overview of the examined use cases. To include the storage device latencies, we also present results for additional storage media.

\begin{figure}[!t]
  \centering
  \includegraphics[width=0.87\linewidth]{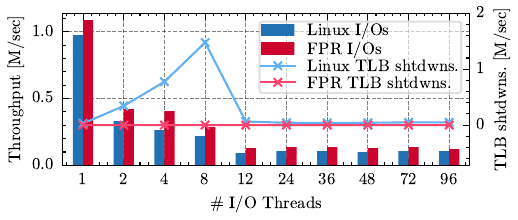}
  \caption{\textit{munmap} Case~1. Y-axis shifted for visibility.}
  \label{fig:eva.mmap-munmap.case1}
\end{figure}

\begin{figure}[!b]
  \centering
  \includegraphics[width=0.87\linewidth]{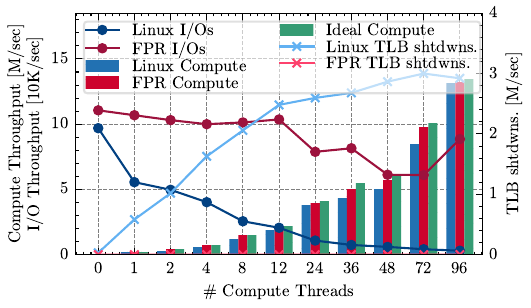}
  \caption{\textit{munmap} Case~2 with 1 I/O and N compute threads.}
  \label{fig:eva.mmap-munmap.case2}
\end{figure}

Fig.~\ref{fig:eva.mmap-munmap.case1} shows the throughput and the number of TLB shootdowns received for Case~1 that has been investigated in previous papers, e.g., based on the vm-scalability benchmark\footnote{https://git.kernel.org/pub/scm/linux/kernel/git/wfg}. 
In the experiments with up to 8 threads, FPR improves performance by up to 30\% compared to the baseline kernel by eliminating nearly all TLB shootdowns. 
When increasing the number of threads, the benchmark triggers a scalability limitation of the kernel's virtual memory management that uses the buddy allocator after nearly every \texttt{munmap} and the performance drops heavily independent of the number of TLB shootdowns (see also \cite{PapagiannisMB21}). 
Addressing this bottleneck is beyond the scope of this paper, whereas FPR will provide higher performance improvements after it has been resolved.

Real applications do not only \texttt{mmap/munmap} pages but also perform compute. 
Case~2, e.g., limits the number of I/O threads to one and adds additional compute threads. 
This case represents applications employing a dedicated thread to perform I/O while other threads process data. 
The I/O thread does not suffer from the scalability limitations mentioned in Case~1.
Fig.~\ref{fig:eva.mmap-munmap.case2} shows the throughput and the number of shootdowns received for a varying number of compute threads. 
The ideal compute baseline does not execute the I/O thread and the compute threads, therefore, do not suffer from shootdowns. 
FPR does not perform any shootdowns and offers a performance very close to the ideal case. 
FPR improves compute performance by up to 21\% compared to the baseline and significantly increases the number of I/O operations. 
This is because the FPR I/O thread does not need to wait for confirmations that the compute threads have finished the shootdowns. 
In contrast, the I/O throughput of the baseline kernel significantly suffers from waiting for these confirmations. Nevertheless, the FPR's I/O drops when crossing NUMA domains at 24 cores and socket domains at 48 cores.  

\begin{figure}[!t]
  \centering
  \includegraphics[width=0.87\linewidth]{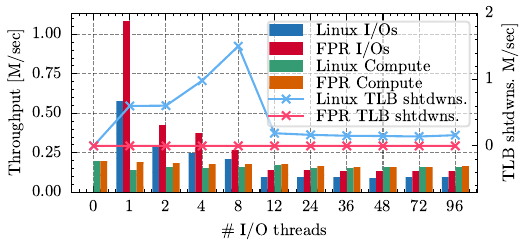}
  \caption{\textit{munmap} Case~3 with 1 compute and N I/O threads.}
  \label{fig:eva.mmap-munmap.case3}
\end{figure}

Varying the number of I/O threads while having a single compute thread (Case~3) performs similarly to Case~1 (see Fig.~\ref{fig:eva.mmap-munmap.case3}). 
For the case that the application contains one I/O and one computation thread, FPR achieves a 93\% higher performance for the I/O thread than the baseline kernel.

\begin{figure}[!b]
  \centering
  \includegraphics[width=0.86\linewidth]{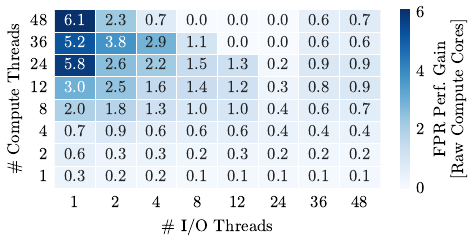}
  \caption{\textit{munmap} Case~4. N I/O, N compute threads.}
  \label{fig:eva.mmap-munmap.case4}
\end{figure}

Case~4 varies the number of I/O and compute threads at the same time. The results for the Intel server have been either skewed towards compute or I/O when the number of threads was higher than the number of cores. We, therefore, used the AMD node for these experiments. Fig.~\ref{fig:eva.mmap-munmap.case4} shows the compute improvements of FPR over the baseline kernel, which are normalized to a single core running computations without any I/O. A value of 6.1, e.g., denotes an improvement equivalent to the compute performance of 6.1 cores compared to the baseline. The highest improvement belongs to cases with a high number of compute threads and a low number of I/O threads. With an increasing number of I/O threads, the improvements degrade similar to Case~1. The number of I/O operations is increased by up to 2.1$\times$ (not shown in Fig.~\ref{fig:eva.mmap-munmap.case4}).

\begin{figure}[!t]
  \centering
  \includegraphics[width=0.87\linewidth]{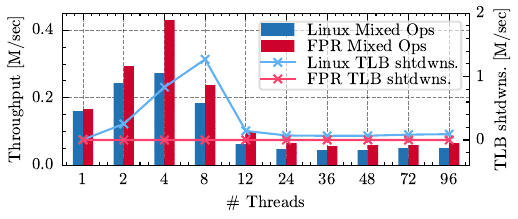}
  \caption{\textit{munmap} Case~5. N mixed threads.}
  \label{fig:eva.mmap-munmap.case5}
\end{figure}

Case~5 mimics applications with mixed threads that perform I/O and compute. Fig.~\ref{fig:eva.mmap-munmap.case5} shows the throughput and the number of received TLB shootdowns for various thread counts. By comparing this figure and Fig.~\ref{fig:eva.mmap-munmap.case1}, we can see that the mixed threads are more affected by shootdowns than I/O threads. This is because mixed threads currently running compute cannot delay the handling of shootdowns until they return from an I/O function (see Figure \ref{fig:TLB-impact-usage-patterns}). 
FPR improves performance by 57\% for 4 threads and still provides 31\% higher performance than the baseline for more than 4 threads, when the VMA starts to bottleneck.

\begin{figure}[!b]
  \centering
  \includegraphics[width=0.87\linewidth]{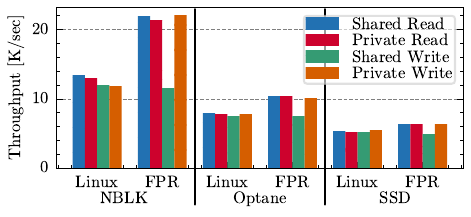}
  \caption{Access modes and devices on Intel server.}
  \label{fig:eva.mmap-munmap.device-comparison}
\end{figure}

\subsubsection{Performance on storage devices}

We re-run Case~2 for various storage devices to understand the impact of their latencies. Fig.~\ref{fig:eva.mmap-munmap.device-comparison} shows the performance for different \texttt{mmap} access modes, performed as random accesses. Shared writes that include a writeback after \texttt{munmap} are not bottlenecked by TLB shootdowns, but by the memory management's write back mechanics, also on the \textit{nullblk}. FPR's focus is mainly on read and private-write accesses. In these cases, FPR improves performance significantly. As expected, the number of I/Os drops compared to running on \textit{nullblk} and TLB shootdowns have a lower impact on performance. Nevertheless, we provide around 18\% performance improvement compared to the baseline kernel for all target cases when using Optane SSDs. Moving to slower SSDs, we see a drop in the speedup based on slower I/Os. 

We also measured the performance on a 10-core server equipped with Optane persistent memory (not shown in  Fig.~\ref{fig:eva.mmap-munmap.device-comparison}). 
FPR provides 38\% and 12\% performance improvement for \textit{nullblk} and persistent memory, respectively on this server. 
Comparing these improvements to those of the Intel 96 core server show that the relative improvement on persistent memory is higher than for Optane SSDs and that our improvements increase for faster storage devices.

\subsubsection{Apache}

\begin{figure}[b!]
  \centering
  \includegraphics[width=0.89\linewidth]{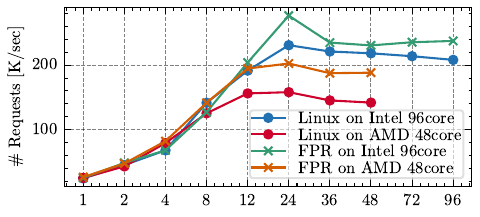}
  \caption{Apache web server}
  \label{fig:eva.apache}
\end{figure}

The Apache web server~\cite{HuNY99} performs \texttt{mmap}-read-\texttt{munmap}s for each request to stream contents from file mappings. 
To evaluate FPR on Apache, we used Apache version \textit{2.4.53} and the \textit{Wrk} workload generator\footnote{HTTP benchmarking tool at \url{ https://github.com/wg/wrk}} accessing the traditional SSDs and \textit{EXT4} . 
Similar to previous works~\cite{Amit17, KumarMKVYKBK18, AmitTW20}, we left Wrk parameters at the default of 6 generator threads using 400 connections.
Apache also runs default settings, using the \texttt{mpm\_event} module that runs multiple server processes and threads. The duration of each test was 30 seconds. Wrk and Apache were pinned to separate cores.

Fig.~\ref{fig:eva.apache} shows the Apache throughput for different core numbers. FPR improves throughput starting at 12 threads and the peak throughput at 24 threads is improved by 22\% and 28\% for the Intel, AMD server, respectively. Scaling further decreases overall throughput whereas FPR is still 6\% and 32\% faster for Intel and AMD servers at 48 cores than the baseline. 

\subsection{Page eviction}
\label{sec:eva:page-eviction}

\begin{figure}[]
  \centering
  \includegraphics[width=0.95\linewidth]{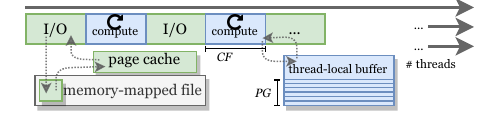}
  \caption{Eviction case overview. }
  \label{fig:eva.page-eviction.overview}
\end{figure}

\begin{figure*}[]
\begin{minipage}[t]{0.309\textwidth}
  \includegraphics[width=\linewidth]{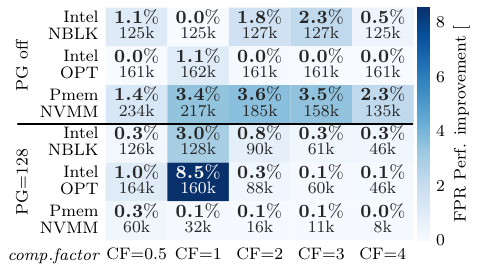}
  \vspace*{-4mm}
  \caption{\textit{Eviction}: FPR improv.}
  \label{fig:eva.page-eviction.heatmap.PGvsnoPG}
\end{minipage}
\hfill
\begin{minipage}[t]{0.329\textwidth}
  \includegraphics[width=\linewidth]{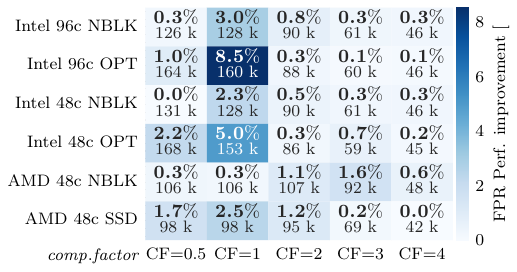}
  \vspace*{-4mm}
  \caption{\textit{Eviction}: Intel vs. AMD.}
  \label{fig:eva.page-eviction.heatmap.INTELvsAMD}
\end{minipage}
\hfill
\begin{minipage}[t]{0.349\textwidth}
  \includegraphics[width=\linewidth]{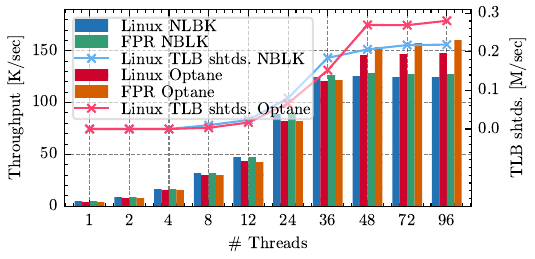}
  \vspace*{-4mm}
  \caption{\textit{Eviction}: Scalability on Intel.}
  \label{fig:eva.page-eviction.scalability}
  \vfill 
\end{minipage}
\end{figure*}

Our page eviction benchmark creates a file larger than the available memory and sets up a \texttt{mmap}. We then spawns threads randomly reading from the mapping. This forces \texttt{kswapd} to evict pages to free space for new pages allocated during page faults. The file size has been set to at least $10\times$ the memory size to reduce the effect of page cache hits and \texttt{madvise} has been set to \texttt{random}. Similar to Case~5 of the previous experiments, we have added compute operations between I/O accesses and scaled the amount of computation by a compute factor \texttt{CF} with 1CF=10K-ops. An optional main memory area for each thread, that is accessed during each compute operation can be configured in multiples of 4 kBytes (\texttt{PG}).

Fig.~\ref{fig:eva.page-eviction.overview} shows how the operations are performed. 
A smaller \texttt{CF} leads to more time spent in the kernel and a higher number of lazy shootdowns, whereas higher \texttt{CF} lowers the  number of I/O operations. 
As the addresses for the \texttt{PG} buffer are not changing, shootdowns might negatively impact access times, because of the page walks forced by emptied TLBs.

Fig.~\ref{fig:eva.page-eviction.heatmap.PGvsnoPG} shows the relative improvement of FPR (top numbers in each cell) and the Linux baseline throughput (bottom numbers). 
We scaled \texttt{CF} from 0.5 to 4.0 and either turned off local buffers (\texttt{PG}-off) or set \texttt{PG} to 128. 
We spawned one thread per available core that reads from a shared \textit{nullblk} mapping and performed the compute phase alternating. 
The benchmark reveals that \texttt{CF} and \texttt{PG} significantly affect performance.
This shows a strong performance potential for saved TLB shootdowns in the eviction path.

The effect of the computation time on our improvements is not linear. It peaks at \texttt{CF}=1 
and then reduces. This experiment reveals that applications performing I/O and compute encounter unexpected performance degradation by shootdowns. The local \texttt{PG} memory accesses during the compute phase also affect our improvements. For the Intel server using an Optane SSD, enabling the \texttt{PG} accesses increases our improvements from 1.1\% to 8.5\%. On the pmem server, removing the memory accesses increases our improvements, because its CPU runs at a higher frequency and the larger number of operations lead to more shootdowns in the baseline. When we enable accesses to the \texttt{PG} buffer, the number of operations and shootdowns decreases because of the slower memory. 

Note that the Optane SSD performs better than the \textit{nullblk} device in this experiment. This anomaly can be partially explained by the lazy shootdown optimization, which prevents shootdowns while the thread is in the kernel and issues the local TLB flush when the thread returns to user-space. The Optane device has a higher latency, and therefore the duration of read or write syscalls is longer than for \textit{nullblk}. Fig.~\ref{fig:eva.page-eviction.scalability}  shows that \textit{nullblk} is faster up to 48 cores, and that then the performance difference increases after adding a second socket.

We performed the experiments on Intel and AMD servers. The AMD server only has 48 cores and we therefore added results for the Intel server for 48 cores with 24 processes pinned to each socket (see Fig.~\ref{fig:eva.page-eviction.heatmap.INTELvsAMD}). Reducing the number of cores, the peak still happens for the same configuration, however, the improvements decrease. The baseline of the AMD server is lower than of the Intel server because of smaller cache sizes and lower frequency. Therefore, the improvement peak is also smaller for the AMD server.

We evaluated FPR also for different access modes in the eviction scenario. The difference between shared and private reads is negligible. For shared writes, FPR improves performance by up to 1.5\%. Private writes are not common at high frequencies, as they require keeping several versions of each page, what would result in out-of-memory errors quickly. 

Our benchmark suite has found different patterns, for which shootdowns have a significant performance impact. Additionally, it enables us to show peaks of the effect of shootdowns when varying the amount and type of user-space processing. 

\subsubsection{Memory mapped databases}

\begin{figure*}
\begin{minipage}[t]{0.155\textwidth}
  \includegraphics[width=\linewidth]{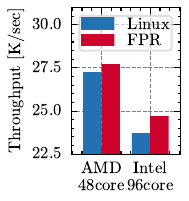}
  \vspace*{-4mm}
  \caption{LMDB.}
  \label{fig:eva.LMDB}
\end{minipage}
\hfill
\begin{minipage}[t]{0.25\textwidth}
  \includegraphics[width=\linewidth]{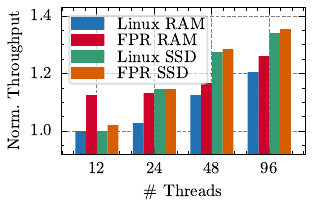}
  \vspace*{-4mm}
  \caption{LevelDB on AMD.}
  \label{fig:eva.LEVELDB-epyc}
\end{minipage}
\hfill
\begin{minipage}[t]{0.305\textwidth}
  \includegraphics[width=\linewidth]{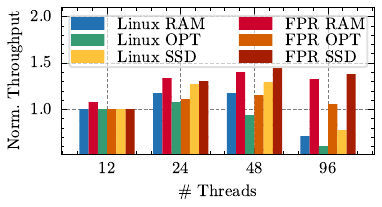}
  \vspace*{-4mm}
  \caption{LevelDB on Intel.}
  \label{fig:eva.LEVELDB-cala}
\end{minipage}
\hfill
\begin{minipage}[t]{0.272\textwidth}
  \includegraphics[width=\linewidth]{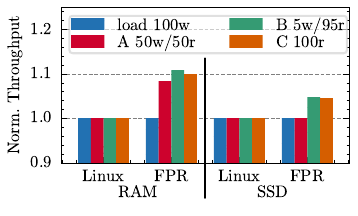}
  \vspace*{-4mm}
  \caption{LevelDB Intel 10 core.}
  \label{fig:eva.LEVELDB-workloads}
\end{minipage}
\end{figure*}

Key-value stores like LMDB~\cite{chu2020lightning} and LevelDB~\cite{ghemawat2014leveldb} use large out-of-bound \texttt{mmap}s. LMDB employs a large and sparse file to store the data and uses \texttt{mmap} to access it. If the file is larger than the memory, \texttt{kswapd} needs to evict pages, causing shootdowns. LevelDB also accesses data through \texttt{mmap} but splits the data into smaller files. Therefore, in addition to page eviction, \texttt{munmaps} are another source of shootdowns. We used YCSB~\cite{CooperSTRS10} to evaluate FPR for these databases. The YCSB workloads involve consistency checks and ran for several hours. First, we selected the read-heavy lookup workload YCSB-C. Fig.~\ref{fig:eva.LMDB} shows the LMDB throughput of the baseline Linux and FPR. The number of threads is set to the number of available CPU cores. FPR provides 1.8\% and 4.0\% performance improvement for Intel and AMD servers. 

Fig.~\ref{fig:eva.LEVELDB-epyc} shows the YCSB-C throughput of LevelDB using the AMD server. All values are normalized to the Linux baseline running with 12 threads on an SSD and a RAM disk. Our improvements are higher for smaller thread numbers because the scalability limitations prevent LevelDB from performing more operations. For 12 threads on the RAM disk, we provide a 12.5\% higher performance than the baseline. Increasing the number of threads and using slower storage devices decreases performance improvements. Note that the 96 threads result uses simultaneous multithreading (SMT) and was done for comparison to the following Intel results.

Fig.~\ref{fig:eva.LEVELDB-cala} shows the corresponding results for the Intel server. 
Increasing the number of threads also increases our improvements. For 48 threads, we provide 20\%, 48\%, and 14\% improvements for the RAM disk, Optane SSD, and traditional SSD, respectively. This difference can originate from either the size of caches and TLBs in the CPUs or from employing different mechanisms for handling shootdowns and reducing their overheads. When using all cores, we observe an unexpected result. The performance of FPR is slightly reduced compared to the previous thread count (48 threads) and the baseline performance also significantly decreases. For 96 threads, the RAM disk under the baseline Linux is showing a performance degradation of 71\% compared to 12 threads. Starting from 48 threads, the Optane SSD also shows performance degradation. Our investigations show that this is based on the internal coherency mechanisms used in the CPUs. However, a full investigation of the reasons is beyond the scope of this paper. Nevertheless, the degradation is highly decreased using FPR by reducing the number of shootdowns to only 2-15\% of the baseline (not shown in the figure).

Fig.~\ref{fig:eva.LEVELDB-workloads} compares different YCSB workloads using LevelDB on the Intel 10-core server. The \emph{load} phase, consisting only of write accesses, shows an unchanged performance. FPR improves the performance of YCSB-A (50\% read, 50\% write) on \textit{nullblk} by 8\%, while the performance on the SSD remains unchanged due to a more bottlenecked write path. The performance of YCSB-B (95\% read, 5\% write) and YCSB-C (100\% read) are improved by 10\% on \textit{nullblk} and 5\% on the SSD.

\subsection{FPR memory and performance overheads}

\begin{figure}
      \centering
      \includegraphics[width=0.8\linewidth]{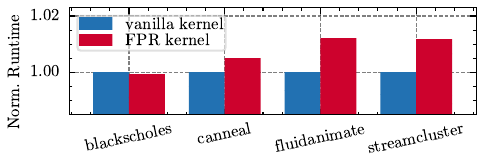}
      \caption{PARSEC overhead.}
      \label{fig:eva.parsec}
\end{figure}

The memory overhead of FPR for its metadata can be calculated to be 0.2\%. To understand the performance overhead for tracking data, we designed two experiments. First, we employed our modified Linux kernel, but without enabling the FPR for the memory mapping. This way, our overheads on \texttt{mmap}-heavy applications that do not use FPR can be measured. FPR imposes at most 1\% processing overhead for all experiments. For real-world applications, the overhead is even lower than for the microbenchmarks (down to unmeasurable for Apache) and becomes nearly negligible.

In the second set of experiments, we selected several applications that do not rely on \texttt{mmap} or other sources of TLB shootdowns. We chose four applications from the PARSEC~\cite{pact/BieniaKSL08} benchmark suite that do not issue any shootdowns. Fig.~\ref{fig:eva.parsec} shows the average of 25 normalized runtimes of the applications. The overhead of FPR depends on the memory allocation pattern of the application. It is between 0\% for \textit{blackscholes} and 1.2\% for \textit{fluidanimate}. Note that many applications allocate the memory only at the beginning of the execution, so that overhead is reduced for longer running applications.

\subsection{Comparison with related work and between kernel versions}

\begin{figure}[]
  \centering
  \includegraphics[width=0.81\linewidth]{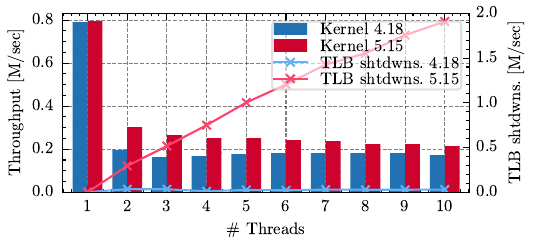}
  \caption{Performance of different kernels for Case~1.
  }
  \label{fig:eva.kernels}
\end{figure}

\red{
Fig.~\ref{fig:eva.kernels} shows Case~1 of the \texttt{munmap} benchmark for baseline Linux 4.18 and 5.15. Kernel 5.15 is more scalable and its performance is less degraded for higher threads counts. When both kernels are limited by the slow page allocator, the 5.15 can even double the performance, showing significant optimizations. Interestingly, the raw number of shootdowns is higher than that of the kernel 4.18, which is explainable by 5.15 not being completely bottlenecked. In summary, we cannot directly compare performance improvements between experiments when the benchmarks are not run on the same kernel version. However, some of the optimizations from previous work like ~\cite{Amit17} are already included in 5.15, so that our improvements are on top of their improvements.
}

\red{
Fastmap compiles at least for a 4.14 kernel \cite{PapagiannisMB21}, running on our 10-core Intel server, and we therefore ported FPR back to it. Fastmap optimizes eviction performance by completely redesigning \texttt{mmap}, including a customized page cache. Unfortunately, Fastmap was not stable in case of frequent munmaps and we were only able to run it for the page eviction microbenchmark from Section \ref{sec:eva:page-eviction}. In this case, Fastmap provides 8\% performance improvement, whereas FPR offers 2\% improvement. The authors of Fastmap claim that 24\% of their improvement is from optimized TLB shootdowns, which is on par with our improvements. In summary, a complete redesign like Fastmap offers a high potential for performance improvement for general mmap at the cost of compatibility, whereas FPR offers immediate performance improvements without breaking the standard Linux design.

We also ported FPR to the most recent kernel 6.10. 
The baseline results without FPR have changed slightly, both for the better and for the worse, depending on the exact scenario.
However, the gain that FPR brought to the baseline has remained the same in terms of quality.
As mentioned before, we decided to show the results of 5.15 as we think it is more valuable, to present a broad varieties of hardware.
}

\section{Conclusion}
\label{sec:conclusion}

TLB shootdowns degrade the performance of many data-intensive applications using \texttt{mmap}ped I/O.
Previous studies have proposed several approaches to mitigate this problem. 
However, they do not address the problem in a standardized way and/or impose overhead on the system. 
By carefully studying the impact of TLB shootdowns, we also uncovered use cases that suffer from this problem that were not identified by previous work. 
Based on our analysis, we propose FPR, a lightweight and transparent solution to mitigate TLB shootdowns. FPR avoids shootdowns when pages are recycling by reoccurring allocations within mmaps. Our experimental results show that FPR improves performance by up to 92\% in microbenchmarks and 28\% in real-world applications. Future work will also investigate approaches to enable FPR to support virtualization\blue{ and includes ports to architectures like \emph{ARM}}.

\bibliographystyle{IEEEtranS}
\bibliography{main.bib}

\begin{thebibliography}{10}
\providecommand{\url}[1]{#1}
\csname url@samestyle\endcsname
\providecommand{\newblock}{\relax}
\providecommand{\bibinfo}[2]{#2}
\providecommand{\BIBentrySTDinterwordspacing}{\spaceskip=0pt\relax}
\providecommand{\BIBentryALTinterwordstretchfactor}{4}
\providecommand{\BIBentryALTinterwordspacing}{\spaceskip=\fontdimen2\font plus
\BIBentryALTinterwordstretchfactor\fontdimen3\font minus
  \fontdimen4\font\relax}
\providecommand{\BIBforeignlanguage}[2]{{%
\expandafter\ifx\csname l@#1\endcsname\relax
\typeout{** WARNING: IEEEtranS.bst: No hyphenation pattern has been}%
\typeout{** loaded for the language `#1'. Using the pattern for}%
\typeout{** the default language instead.}%
\else
\language=\csname l@#1\endcsname
\fi
#2}}
\providecommand{\BIBdecl}{\relax}
\BIBdecl

\bibitem{Amit17}
N.~Amit, ``Optimizing the {TLB} shootdown algorithm with page access
  tracking,'' in \emph{2017 {USENIX} Annual Technical Conference (ATC), Santa
  Clara, CA, USA, July 12-14, 2017}, 2017, pp. 27--39.

\bibitem{AmitTW20}
N.~Amit, A.~Tai, and M.~Wei, ``Don't shoot down {TLB} shootdowns!'' in
  \emph{Fifteenth EuroSys Conference ({EuroSys}), Heraklion, Greece, April
  27-30}, 2020, pp. 35:1--35:14.

\bibitem{AwadBBSL17}
A.~Awad, A.~Basu, S.~Blagodurov, Y.~Solihin, and G.~H. Loh, ``Avoiding {TLB}
  shootdowns through self-invalidating {TLB} entries,'' in \emph{26th
  International Conference on Parallel Architectures and Compilation Techniques
  (PACT), Portland, OR, USA, September 9-13}, 2017, pp. 273--287.

\bibitem{pact/BieniaKSL08}
C.~Bienia, S.~Kumar, J.~P. Singh, and K.~Li, ``The {PARSEC} benchmark suite:
  characterization and architectural implications,'' in \emph{17th
  International Conference on Parallel Architectures and Compilation Techniques
  ({PACT}), Toronto, Ontario, Canada, October 25-29}, 2008, pp. 72--81.

\bibitem{ChangK97}
M.-S. Chang and K.~Koh, ``Lazy tlb consistency for large-scale
  multiprocessors,'' in \emph{Proceedings of IEEE International Symposium on
  Parallel Algorithms Architecture Synthesis, Aizu-Wakamtsu, Japan, March
  17-21}, 1997, pp. 308--315.

\bibitem{chu2020lightning}
H.~Chu, ``Lightning memory-mapped database manager {(LMDB)},'' 2020.

\bibitem{CooperSTRS10}
B.~F. Cooper, A.~Silberstein, E.~Tam, R.~Ramakrishnan, and R.~Sears,
  ``Benchmarking cloud serving systems with {YCSB},'' in \emph{Proceedings of
  the 1st {ACM} Symposium on Cloud Computing (SoCC), Indianapolis, Indiana,
  USA, June 10-11, 2010}, 2010, pp. 143--154.

\bibitem{CrottyLP22}
A.~Crotty, V.~Leis, and A.~Pavlo, ``Are you sure you want to use {MMAP} in your
  database management system?'' in \emph{12th Conference on Innovative Data
  Systems Research (CIDR), Chaminade, CA, USA, January 9-12}, 2022.

\bibitem{DechevPS10}
D.~Dechev, P.~Pirkelbauer, and B.~Stroustrup, ``Understanding and effectively
  preventing the {ABA} problem in descriptor-based lock-free designs,'' in
  \emph{13th {IEEE} International Symposium on
  Object/Component/Service-Oriented Real-Time Distributed Computing (ISORC),
  Carmona, Sevilla, Spain, 5-6 May}, 2010, pp. 185--192.

\bibitem{EssenPAG12}
B.~V. Essen, R.~A. Pearce, S.~Ames, and M.~B. Gokhale, ``On the role of {NVRAM}
  in data-intensive architectures: An evaluation,'' in \emph{26th {IEEE}
  International Parallel and Distributed Processing Symposium (IPDPS),
  Shanghai, China, May 21-25}, 2012, pp. 703--714.

\bibitem{EversBC22}
M.~Evers, L.~Barnes, and M.~Clark, ``The {AMD} next-generation "zen 3" core,''
  \emph{{IEEE} Micro}, vol.~42, no.~3, pp. 7--12, 2022.

\bibitem{ghemawat2014leveldb}
S.~Ghemawat and J.~Dean, ``{LevelDB}, a fast and lightweight key/value database
  library by {Google},'' 2014.

\bibitem{GosakanHKMMSTWB23}
K.~Gosakan, J.~Han, W.~Kuszmaul, I.~N. Mubarek, N.~Mukherjee, K.~Sriram,
  G.~Tagliavini, E.~West, M.~A. Bender, A.~Bhattacharjee, A.~Conway,
  M.~Farach{-}Colton, J.~Gandhi, R.~Johnson, S.~Kannan, and D.~E. Porter,
  ``Mosaic pages: Big {TLB} reach with small pages,'' in \emph{Proceedings of
  the 28th {ACM} International Conference on Architectural Support for
  Programming Languages and Operating Systems, Volume 3, {ASPLOS} 2023,
  Vancouver, BC, Canada, March 25-29, 2023}, T.~M. Aamodt, N.~D.~E. Jerger, and
  M.~M. Swift, Eds.\hskip 1em plus 0.5em minus 0.4em\relax {ACM}, 2023, pp.
  433--448.

\bibitem{GugaleGMJ20}
H.~Gugale, N.~Gulur, Y.~Marathe, and L.~K. John, ``{ATTC} (@c): Addressable-tlb
  based translation coherence,'' in \emph{International Conference on Parallel
  Architectures and Compilation Techniques (PACT), Virtual Event, GA, USA,
  October 3-7}, 2020, pp. 481--492.

\bibitem{HuNY99}
Y.~Hu, A.~K. Nanda, and Q.~Yang, ``Measurement, analysis and performance
  improvement of the apache web server,'' in \emph{Proceedings of the {IEEE}
  International Performance Computing and Communications Conference (IPCCC),
  Phoenix/Scottsdale, Arizona, USA, 10-12 February}, 1999, pp. 261--267.

\bibitem{IntelManual}
Intel, \emph{Intel 64 and IA-32 Architectures Software Developer’s Manual
  Volume 2 (2A, 2B, 2C, \& 2D): Instruction Set Reference, A-Z}, 2023.

\bibitem{Knowlton65}
K.~C. Knowlton, ``A fast storage allocator,'' \emph{Commun. {ACM}}, vol.~8,
  no.~10, pp. 623--624, 1965.

\bibitem{KumarMKVYKBK18}
M.~Kumar, S.~Maass, S.~Kashyap, J.~Vesel{\'{y}}, Z.~Yan, T.~Kim,
  A.~Bhattacharjee, and T.~Krishna, ``{LATR:} lazy translation coherence,'' in
  \emph{Proceedings of the Twenty-Third International Conference on
  Architectural Support for Programming Languages and Operating Systems
  ({ASPLOS}), Williamsburg, VA, USA, March 24-28}, 2018, pp. 651--664.

\bibitem{MaassKKKB20}
S.~Maass, M.~K. Kumar, T.~Kim, T.~Krishna, and A.~Bhattacharjee, ``{ECOTLB:}
  eventually consistent tlbs,'' \emph{{ACM} Trans. Archit. Code Optim.},
  vol.~17, no.~4, pp. 27:1--27:24, 2020.

\bibitem{PapagiannisMB21}
A.~Papagiannis, M.~Marazakis, and A.~Bilas, ``Memory-mapped {I/O} on
  steroids,'' in \emph{EuroSys '21: Sixteenth European Conference on Computer
  Systems, Online Event, United Kingdom, April 26-28, 2021}, 2021, pp.
  277--293.

\bibitem{PapagiannisXSMB20}
A.~Papagiannis, G.~Xanthakis, G.~Saloustros, M.~Marazakis, and A.~Bilas,
  ``Optimizing memory-mapped {I/O} for fast storage devices,'' in \emph{2020
  {USENIX} Annual Technical Conference (ATC), July 15-17}, 2020, pp. 813--827.

\bibitem{ParkMYS19}
J.~Park, C.~Min, H.~Y. Yeom, and Y.~Son, ``z-read: Towards efficient and
  transparent zero-copy read,'' in \emph{12th {IEEE} International Conference
  on Cloud Computing ({CLOUD}), Milan, Italy, July 8-13}, 2019, pp. 367--371.

\bibitem{RadojkovicCVPGCNV08}
P.~Radojkovic, V.~Cakarevic, J.~Verd{\'{u}}, A.~Pajuelo, R.~Gioiosa, F.~J.
  Cazorla, M.~Nemirovsky, and M.~Valero, ``Measuring operating system overhead
  on {CMT} processors,'' in \emph{20th International Symposium on Computer
  Architecture and High Performance Computing, {SBAC-PAD} 2008, October 29 -
  November 1, 2008, Campo Grande, MS, Brazil}, E.~N. C{\'{a}}cares, W.~Cirne,
  and V.~K. Prasanna, Eds.\hskip 1em plus 0.5em minus 0.4em\relax {IEEE}
  Computer Society, 2008, pp. 133--140.

\bibitem{RomanescuLSB10}
B.~F. Romanescu, A.~R. Lebeck, D.~J. Sorin, and A.~Bracy, ``Unified
  instruction/translation/data {(UNITD)} coherence: One protocol to rule them
  all,'' in \emph{16th International Conference on High-Performance Computer
  Architecture {(HPCA}), Bangalore, India, 9-14 January}, 2010, pp. 1--12.

\bibitem{RyooJB18}
J.~H. Ryoo, L.~K. John, and A.~Basu, ``A case for granularity aware page
  migration,'' in \emph{Proceedings of the 32nd International Conference on
  Supercomputing (ICS), Beijing, China, June 12-15}, 2018, pp. 352--362.

\bibitem{SkarlatosDGKT21}
D.~Skarlatos, U.~Darbaz, B.~Gopireddy, N.~S. Kim, and J.~Torrellas,
  ``Babelfish: Fusing address translations for containers,'' \emph{{IEEE}
  Micro}, vol.~41, no.~3, pp. 57--62, 2021.

\bibitem{StamlerHRZP22}
T.~Stamler, D.~Hwang, A.~Raybuck, W.~Zhang, and S.~Peter, ``zio: Accelerating
  io-intensive applications with transparent zero-copy {IO},'' in \emph{16th
  {USENIX} Symposium on Operating Systems Design and Implementation (OSDI),
  Carlsbad, CA, USA, July 11-13}, 2022, pp. 431--445.

\bibitem{Uhlig05}
\BIBentryALTinterwordspacing
V.~Uhlig, ``Scalability of microkernel-based systems,'' Ph.D. dissertation,
  Karlsruhe Institute of Technology, Germany, 2005. [Online]. Available:
  \url{http://digbib.ubka.uni-karlsruhe.de/volltexte/1000004020}
\BIBentrySTDinterwordspacing

\bibitem{Uhlig07}
V.~Uhlig, ``The mechanics of in-kernel synchronization for a scalable
  microkernel,'' \emph{{ACM} {SIGOPS} Oper. Syst. Rev.}, vol.~41, no.~4, pp.
  49--58, 2007.

\bibitem{VillaviejaKVERMNCU11}
C.~Villavieja, V.~Karakostas, L.~Vilanova, Y.~Etsion, A.~Ram{\'{\i}}rez,
  A.~Mendelson, N.~Navarro, A.~Cristal, and O.~S. Unsal, ``Didi: Mitigating the
  performance impact of {TLB} shootdowns using a shared {TLB} directory,'' in
  \emph{International Conference on Parallel Architectures and Compilation
  Techniques (PACT), Galveston, TX, USA, October 10-14}, 2011, pp. 340--349.

\bibitem{WentzlaffA09}
D.~Wentzlaff and A.~Agarwal, ``Factored operating systems (fos): the case for a
  scalable operating system for multicores,'' \emph{{ACM} {SIGOPS} Oper. Syst.
  Rev.}, vol.~43, no.~2, pp. 76--85, 2009.

\bibitem{YanVCB17}
Z.~Yan, J.~Vesel{\'{y}}, G.~Cox, and A.~Bhattacharjee, ``Hardware translation
  coherence for virtualized systems,'' in \emph{Proceedings of the 44th Annual
  International Symposium on Computer Architecture (ISCA), Toronto, ON, Canada,
  June 24-28}, 2017, pp. 430--443.

\end{thebibliography}

\end{document}